\documentstyle[12pt]{article}
\begin{document}

\begin{center}
{\large \bf The contribution of the hadronic  component of the photon
 into the  $D^*$-meson photoproduction at HERA.}

\vspace{2cm}
 Berezhnoy~A.V.

{\it Skobeltsin Institute for Nuclear Research of Moscow State University,

Moscow, Russia}

 Likhoded~A.K.

{\it Institute for High Energy Physics, Protvino, Russia}
    
\vspace{2cm}

\parbox{15cm}{
In the framework of the vector dominance model (VDM) 
new contribution into the cross-section
of $D^*$-meson photoproduction at HERA has been estimated.
This contribution is due to the
interaction of the virtual vector $c\bar c$-mesons
from photon with the proton.
It has been shown that the mechanism under discussion plays
the essential role for
all values of transverse momentum of  $D^*$-mesons at HERA.
Taking into
account for the contribution of this mechanism  improves
the description of the experimental cross section distribution.
  }
\end{center}

\newpage

\newcommand{\xg}{x_{\gamma}^{\rm OBS}}

\section{Introduction}

One of the most interest features of the charm photoproduction at HERA,
which had been   discovered by ZEUS Collaboration, is the
 correlation between jet energies in the  events  with  the
$D^*$-meson \cite{ZEUS}.
  
It was shown experimentally that in the kinematical region $130<W<280$~GeV,
$Q^2<1 \ \rm GeV^2$ (where $W$ is the energy of the $\gamma p$-interaction,
and $Q^2$ is the photon virtuality) in $\sim 30$\% of all events with
$D^*$-mesons two jets with the highest transverse energy take away only a half 
of the photon energy. The distribution over $\xg$ has been used for the
quantitative analysis of this phenomenon. The value of $\xg$
means the part of
the photon energy which is taken away by two jets with the highest transverse
energy. It is defined as below:

\begin{equation}
\xg=\frac{\sum_{\rm jet=1,2}E_T^{\rm jet}e^{-\eta_{\rm jet}}}{2E_ey},
\end{equation}
where the $\eta^{\rm jet}$ -- pseudorapidity of the jet, $E_T^{\rm jet}$--
transverse energy of the jet. One takes the sum over two jets with the highest 
$E_T^{\rm jet}$.

The following equation allows to clarify the physical meaning of $\xg$:

\begin{equation}
2E_ey=\sum_{\rm jet}^{\rm  all}(E^{\rm  jet}-p^{\rm  jet}_z)=\sum_{\rm
jet}^{\rm all} E^{\rm jet}_T\cdot e^{-\eta_{\rm jet}}.
\end{equation}
One can  see  that for events with two jets $\xg \equiv 1$.
For events with many jets always $\xg<1$.

In the framework of the pQCD there are two types of contribution
into charm photoproduction:

1) contribution from the scattering of the parton from the hadron on the
photon  ("direct photon");

2) contribution from  the scattering of the parton from hadron on the
parton from photon ("resolved photon").

For the "direct photon" events $\xg \sim 1$. This effect can be easily
understood if one  keeps in mind that in the LO such events have only two jets.

In the "resolved photon" events or, by other words
in the interaction of the hadronic parton with the charm quark from photon, 
the  charm quark carries only a part of the photon energy,
and that is why $\xg$ can be essentially less than 1.

In the NLO approach one can not distinguish between the contributions
from "direct photon" and "resolved photon"
on the level of the cross-sections,
and NLO matrix element includes  both of these
contributions \cite{Frixione,Kniehl}.

Nevertheless, these contributions have a different kinematical
behavior and can
be distinguished one from another by using the distribution over $\xg$.  
So, ZEUS collaboration divides the $\xg$ range into two parts:

1) $0<\xg<0.75$ ("small" $\xg$) which corresponds to the "resolved photon" events;

2) $0.75<\xg<1$ ("large" $\xg$) which corresponds to the "direct photon" events.

As it follows from the ZEUS  data, for the kinematical region
$p_T^{D^*}>3$~GeV, $|\eta^{D^*}|<1.5$, $130<W<280$~GeV, $Q^2<1\ \rm GeV^2$,
$|\eta^{\rm jet}|<2.4$, $E_T^{\rm jet_1}>7$~GeV, $E_T^{\rm jet_2}>6$~GeV
a noticeable discrepancy between
the NLO predictions for the  distribution over $\xg$
and the experimental one exits \cite{Frixione, BKL}.
The experimental value of the cross-section in the region $0<\xg< 1$
(region of "resolved photon") is  larger than that predicted by
the NLO approach.
 
 In this paper we will try to describe the distribution over $\xg$ by taking into account
the hadronic component of the photon in the framework   of
the Vector Dominance Model (VDM) \cite{VDM,JpsiN}.

 \section{Description of the model and  results}

First of all, it is worth to mention that we have to take into account
only photon charm quark fluctuation, because we are interested
in the events with the charm jets only. In our calculation
we suggest that the main contribution
into the $D^*$-meson photoproduction  at HERA in the region of "small"
$\xg$  is due to the scattering of gluon on the $c$-quark from the $J/\psi$,
$\psi'$ and other vector $c\bar c$-mesons from photon in the
wavefunction of the initial photon.

In the  case under consideration the photon wave function can be approximated 
by a sum of the wave functions
of the vector mesons as bellow \cite{VDM,JpsiN}:

\begin{equation}
|\gamma \rangle
=\sum_{V=\psi,\psi',\psi(3770)...}\frac{4\pi\alpha}{\gamma^2_V} |V\rangle,
\end{equation}
where a $\gamma_V$ constant is determined by the vector meson width of
decay into
 the $e^+e^-$-pair:
\begin{equation}
\Gamma (V \to e^+e^- )=
\frac{\alpha^2}{3}\frac{4\pi}{\gamma^2_V}M_V
\end{equation}

From the equations above
the charm photoproduction cross section
can be expressed through cross-sections of the vector
meson interaction with hadron:

\begin{equation}
\sigma(\gamma N   \to   c  \bar  c)=
\sum_{V=\psi,\psi',\psi(3770)...}  \frac{4\pi\alpha}{\gamma^2_V}
\sigma_{\rm inel.} (VN)
\end{equation}

As it was mentioned above,  one should take  into
account only the $c\bar c$-mesons contribution, as the contribution
from other vector mesons  is small \cite{JpsiN}.

So, in the framework of VDM, the following equation can be written for the
charm photoproduction with two large transverse energy jets:
\begin{equation}
\sigma(\gamma p)=
\sum_{V=\psi,\psi',\psi(3770)...}\frac{4\pi\alpha}{\gamma_V^2}
K(E_{\gamma},m_V,m_p)\cdot \sigma(V p),
\end{equation}
where

\begin{equation}
\sigma(V p)=\int \int f_c^{V}(\xg) f^N_g(x_g)
\sigma(cg \to {\rm 2\ jet_{{\rm higt\ E_T}}}) d\xg dx_g,
\end{equation}
and

\begin{equation}
K(E_{\gamma},m_V,m_p)=\sqrt{\frac{E_V^2-m_p^2}{E_{\gamma}^2}}.
\end{equation}

To evaluate the $\sigma(cg \to {\rm 2\ jet_{{\rm higt\ E_T}}})$ we use
the Born
approximation for the $cg\to cg$ elastic scattering.
So, the cross-section for the events with two high $E_T$ jets
is determined by the hard scattering of the gluon on the $c$-quark
from vector meson, as it has been shown in Fig.~1.
     
For the $p_T$-distribution of $D^*$ we  convolute this cross section with
the fragmentation   function   of   $c$-quark  into  $D^*$-meson  from
\cite{BKL}.

For the $c$-quark distribution in $J/\psi$ and any other
vector  $c\bar c$-meson
we use the Regge parametrization \cite{SFun}:

\begin{equation}
f_c^{\Psi}(x)=N x^{-\alpha_{\Psi}}(1-x)^{\gamma -\alpha_{\Psi}},
\end{equation}
where $x$ is the $J/\Psi$ momentum fraction of $c$-quark,
$\alpha_{\psi}$ is the intercept of $J/\psi$,
$N$ is the normalization
coefficient and $\gamma=1/4$.

The best description of the charm production
at small $p_T$ can be achieved for
$\alpha_{\psi}=-2.2$ \cite{Tolstenkov}.
With this value of $\alpha_{\psi}$ the parametrization
has the following form

\begin{equation}
f_c^{\Psi}(x)=49.8x^{2.2}(1-x)^{2.45}.
\end{equation}

It is worth to mention that this quark distribution
is a complete analog of the
valence quark distribution in $\rho$-meson:
\begin{equation}
f_q^{\rho}(x)=N x^{-\alpha_{\rho}}(1-x)^{\gamma -\alpha_{\rho}}.
\end{equation}

The only difference is the different  values
for Regge intercept and parameter $\gamma$.

Average fraction of the total momentum carred by $c$-quarks
is close to 1:
\begin{equation}
\langle x_c \rangle + \langle x_{\bar c} \rangle \approx 0.96,
\end{equation}
and each quark get approximately a half of the photon momentum.

So, we get the result we needed: in average, only a half of the photon
momentum participates in the hard $cg$-interaction followed by
the production of two jets with large transverse energies.  
Namely such events give the contribution into region of "small" $\xg$.

The characteristic time of the $c\bar c$-fluctuation
is about $\sim 1/m_{\psi}$
in the rest frame of fluctuation. This value multiplied by the Lorence factor
$E_{\gamma}/m_{\psi}$
is essentially larger than the characteristic time of the hard jet production:

\begin{equation}
\frac{E_{\gamma}}{m_{\Psi}^2} \gg \frac{1}{E_T^{\rm jet}}.
\end{equation}
This circumstance allows us to calculate the charm production cross-section
incoherentely, summarizing the contributions
from hard scattering on each $c$-quark.
It is clear that for small $E_T$ such approach is not valid and it is worth
to consider the $c\bar c$-meson as a color dipole and take
into account transverse quark motion and interference between different
contributions. In this kinematical region the dipole model
\cite{Dipole} gives a good description of the data.

In this paper we limited ourselves
to the consideration of the incoherent contribution
 and describe  the  distributions
in $p_T$ and $\eta$ for  $p_T>4$~GeV.

The  cuts  on  the jets transverse energy for the distribution in $\xg$
($E_T^{\rm jet_1}>7$~GeV and $E_T^{\rm jet_2}>6$~GeV) ensure large transverse
momentum of $c$-quark before its fragmentation into $D^*$-meson
(or in other words after hard scattering $cg\to cg$).
That is why we use our model for all investigated
values of  $p_T$  to  predict  the $\xg$ distribution.

Our model does not contain any new parameters. Indeed:
    
1)  the photon coupling constants with
$J/\psi$ and other vector $c\bar c$-mesons
are known from the width of meson decay into $e^+e^-$;

2) the $c$-quark distribution in
the vector $c\bar c$-meson (9)  contains
the Regge intercept $\alpha_{\psi}(0)$ which
determines fragmentation function of $c$-quark into $D^*$,
and value of $\alpha_{\psi}(0)=-2.2$ allows to describe
the experimental data on $c$-quark fragmentation
into $D^*$-meson  and charm  photoproduction  at  low energies
\cite{SFun};

3) the scale $\mu_R$ in the determination of the strong coupling constant
$\alpha_s(\mu_R)$ in the hard scattering matrix element
for the process $cg\to cg$ and the scale $\mu_F$ for the gluonic structure
function are common parameters for such calculations.
We use the following values for the scales: $\mu_R=\mu_F=2m_{D^*}$.

It is worth to mention that the main 
contribution into the charm photoproduction 
calculated in the framework of VDM is due to $J/\psi$ (about 60\%). 
Other vector
$c\bar c$-mesons give   40\% of the total cross-section.

In the frame of the model under consideration the ZEUS experimental 
cuts $E^{jet_1}_T>6$~GeV, $E^{jet_2}_T>7$~GeV are equivalent to the following
ones: $E^{jet_1}_T,E^{jet_2}_T>7$~GeV. The model does
not account for the initial transverse momentum of the $c$-quark into the $c\bar c$-fluctuation, which is about $\sim 1$~GeV. That is why the both high energy jets have the same transverse energy.  

To evaluate the cross-section uncertainty which is due to the transverse momentum of the initial $c$-quark we have calculated the distribution in
$x^{OBS}_{\gamma}$ for $E^{jet_1}_T,E^{jet_2}_T>7$~GeV  and
for $E^{jet_1}_T,E^{jet_2}_T>6$~GeV. The difference between two this
calculations estimates the theory uncertainty.

One can see from Fig.~2a,b that this uncertainty is rather large.

As one can see from Fig.~2a  the description of the experimental
data on  $\xg$-distribution in the region of the "resolved photon"
has been essentially improved by adding
the VDM predictions with the BKL ones \cite{BKL}.

The slightly better
description of the data can be achieved by adding the VDM predictions
to NLO ones from \cite{Frixione} (see Fig.~2b).
However, one must keep in mind that the adding of the VDM contribution can 
dramatically enlarge the normalization of the distribution in transverse momentum of the $D^*$-meson for the model \cite{Frixione}.  
 
It is important to mention that
 hard interaction with the hadronic component of the photon ($J/\psi$,
$\psi'$ and so on) gives
the noticeable  contribution into the region of  large $p_T$ (see Fig.~3).
 In other words, the attempt to describe
 $\xg$ distribution in the framework of VDM
 leads  to additional
contribution in the region of large $p_T$.

So, the following conclusion can be drawn:
the nonperturbative $c\bar c$-fluctuations of photon are important
for the charm photoproduction.  Such fluctuations are described in the
framework of VDM and  can
not be described by simple formula of the pQCD
\begin{equation}
f_c(x)\sim (x^2+(1-x)^2)
\end{equation}
or improved formula \cite{Ryskin}:
\begin{equation}
f_c(x) \sim \left (x^2+(1-x)^2+\frac{2m_c^2}{E_T^2}x(1-x)
\right )
\end{equation}
or Bethe-Heitler formula \cite{BH}:
\begin{equation}
\begin{array}{rcl}
f_c(x)&=&\frac{4\alpha}{3\pi} \left [ \beta \left ( 8x(1-x) -1
-\frac{4m_c^2}{Q^2}x(1-x)\right ) \right.\\
& & \left. + \left ( x^2+(1-x)^2+
\frac{4m_c^2}{Q^2}x(1-3x)-\frac{8m_c^4}{Q^4}x^2 \right )
\ln \left ( \frac{1+\beta}{1-\beta} \right ) \right ],
\end{array}
\end{equation}
if $\beta^2=1-\frac{4m_c^2x}{(1-x)Q^2}>0$  and  $f_c(x)=0$
if $\beta^2<0$.

This part of the fluctuations has a typical hadronic structure (9) and
such fluctuations provide about 30\% in the cross-section distribution
in $\xg$ for the $D^*$-meson photoproduction.

The topology of the events in the region
of the "small" $\xg$  essentially differs from
 one of the events in the region
of the "large" $\xg$.  In the latter case both jets
carry a charm quark, in contradiction
with the former case, where only one jet  has
charm quark. This circumstance leads  to
the large  azimuthal correlation  between charmed
particles  for the "direct photon" and
to negligible one for "resolved photon" (see Fig.~4).

\section{Conclusion} 

It has been shown that scattering of virtual
vector $c\bar c$-mesons from photon on
the proton gives noticeable contribution into
the $D^*$-meson photoproduction  for all values of the transverse
momentum investigated  at HERA.
Furthermore this contribution allows essentially  improve the description
of the experimental distribution in $\xg$ in the region of the "resolved
photon" ($0<\xg<0.75$).

We are grateful L.~Gladilin, V~Kiselev, D.~Kharzeev, I.~Korzhavina
for the useful discussion of the
materials presented in this article.

\newpage
\vspace*{-3cm}\includegraphics{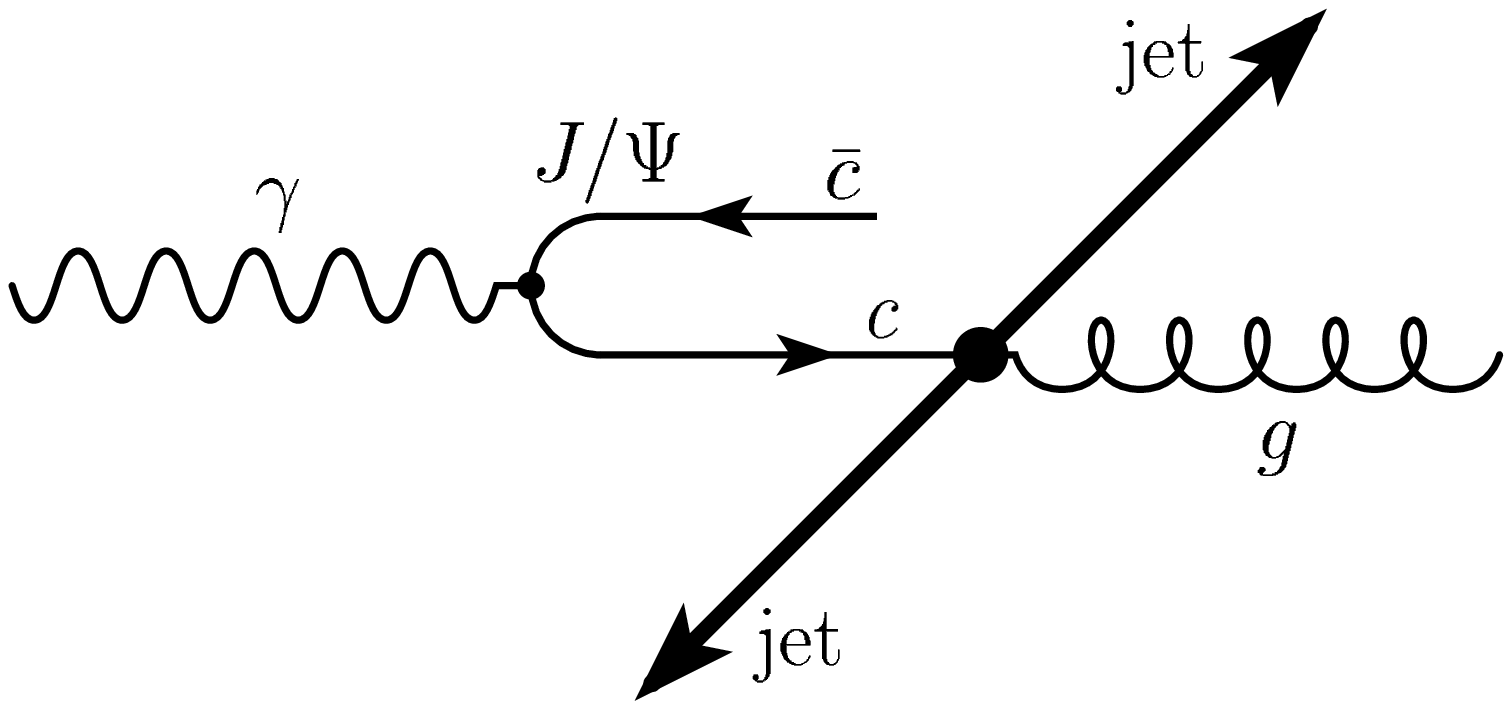}
\begin{picture}(450,450)\put(20,0){Fig.~1. \parbox[t]{12cm}{
Charm production scheme in the Vector Dominance Model.
}}\end{picture}

\newpage\vspace*{-3cm}\includegraphics{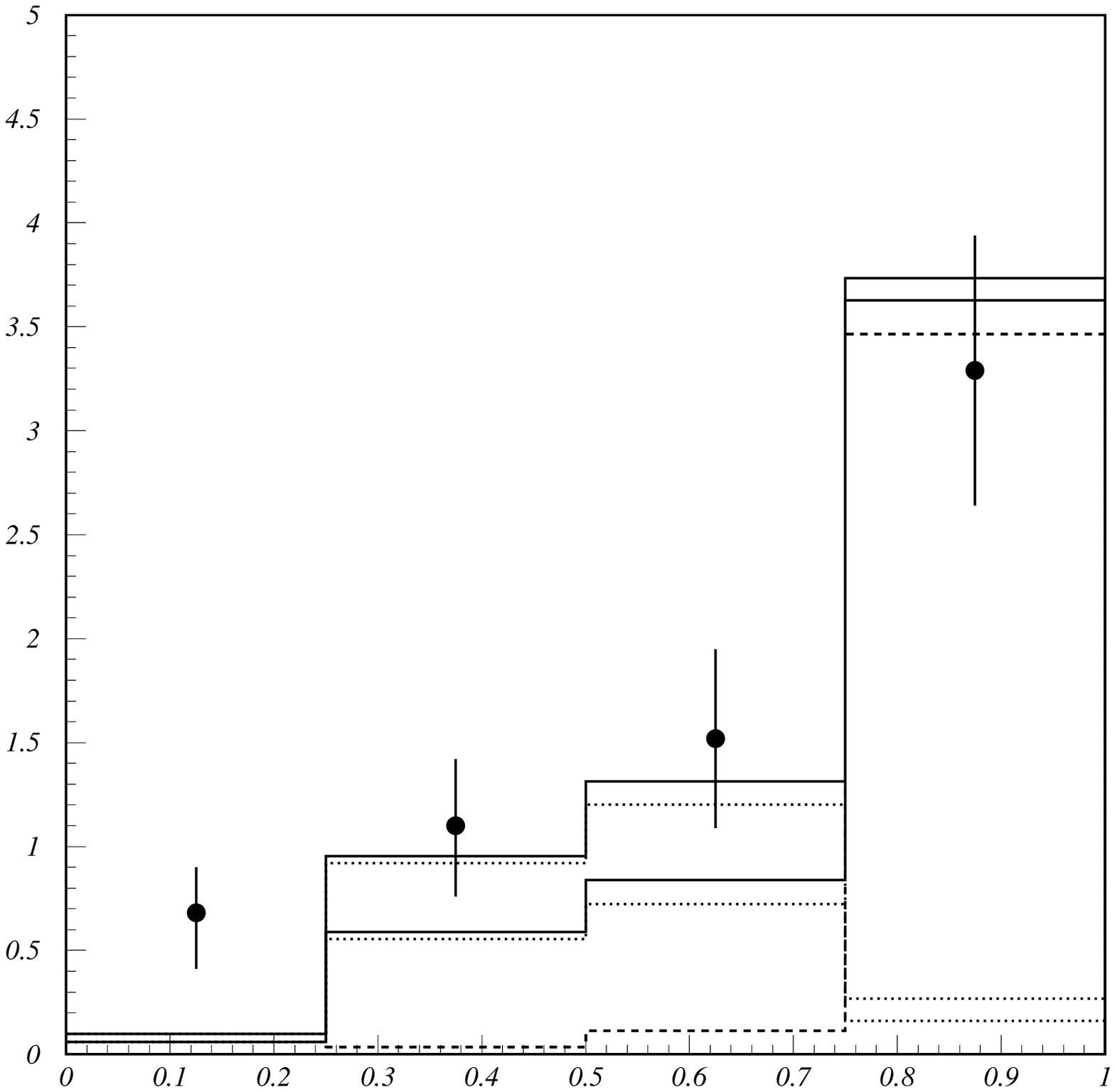}
\begin{picture}(450,450)\put(0,250){$d\sigma_{\gamma p}/d\xg$, nb}
\put(360,-130){$\xg$}
\put(20,-180){Fig.~2a. \parbox[t]{12cm}{
The experimental distribution in $\xg$ of the $D^*$-meson production cross
section (ZEUS) in comparison with the BKL  predictions  \cite{BKL}  (dashed
histograms), with VDM predictions (the region between the dotted histograms) and with the sum of BKL and VDM predictions ( the region between the solid histograms).
}}\end{picture}
\newpage\vspace*{-3cm}\includegraphics{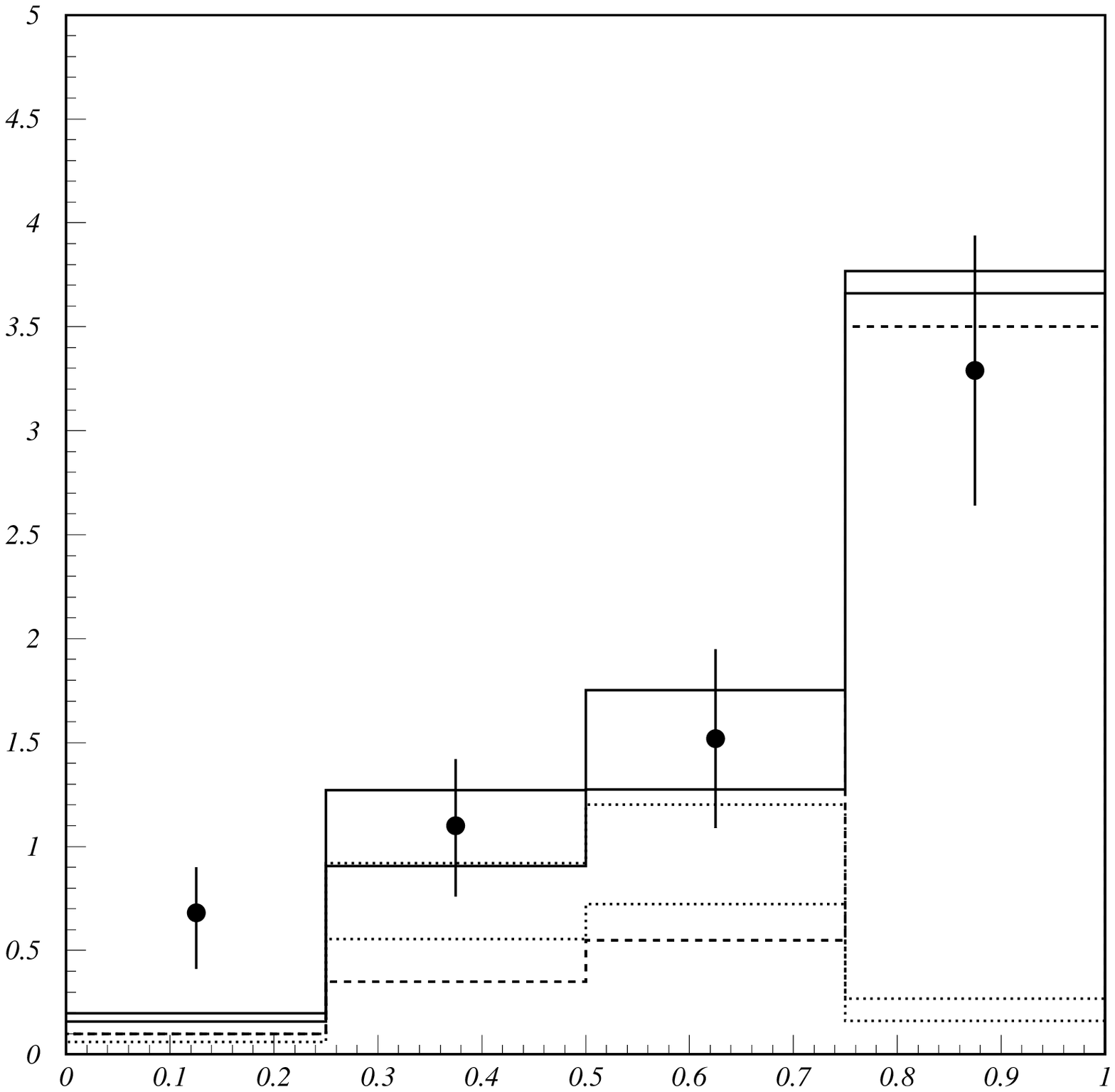}
\begin{picture}(450,450)\put(0,250){$d\sigma_{\gamma p}/d\xg$, nb}
\put(360,-130){$\xg$}
\put(20,-180){Fig.~2b. \parbox[t]{12cm}{
The experimental distribution in $\xg$ of the $D^*$-meson production cross
section (ZEUS) in comparison with the Frixione et al. predictions
\cite{Frixione}  (dashed histogram), with VDM predictions
(the region between the dotted histograms) and with the
sum of Frixione at al. and VDM predictions (the region between the solid histograms).

}}\end{picture}
\newpage\vspace*{-3 cm}\includegraphics{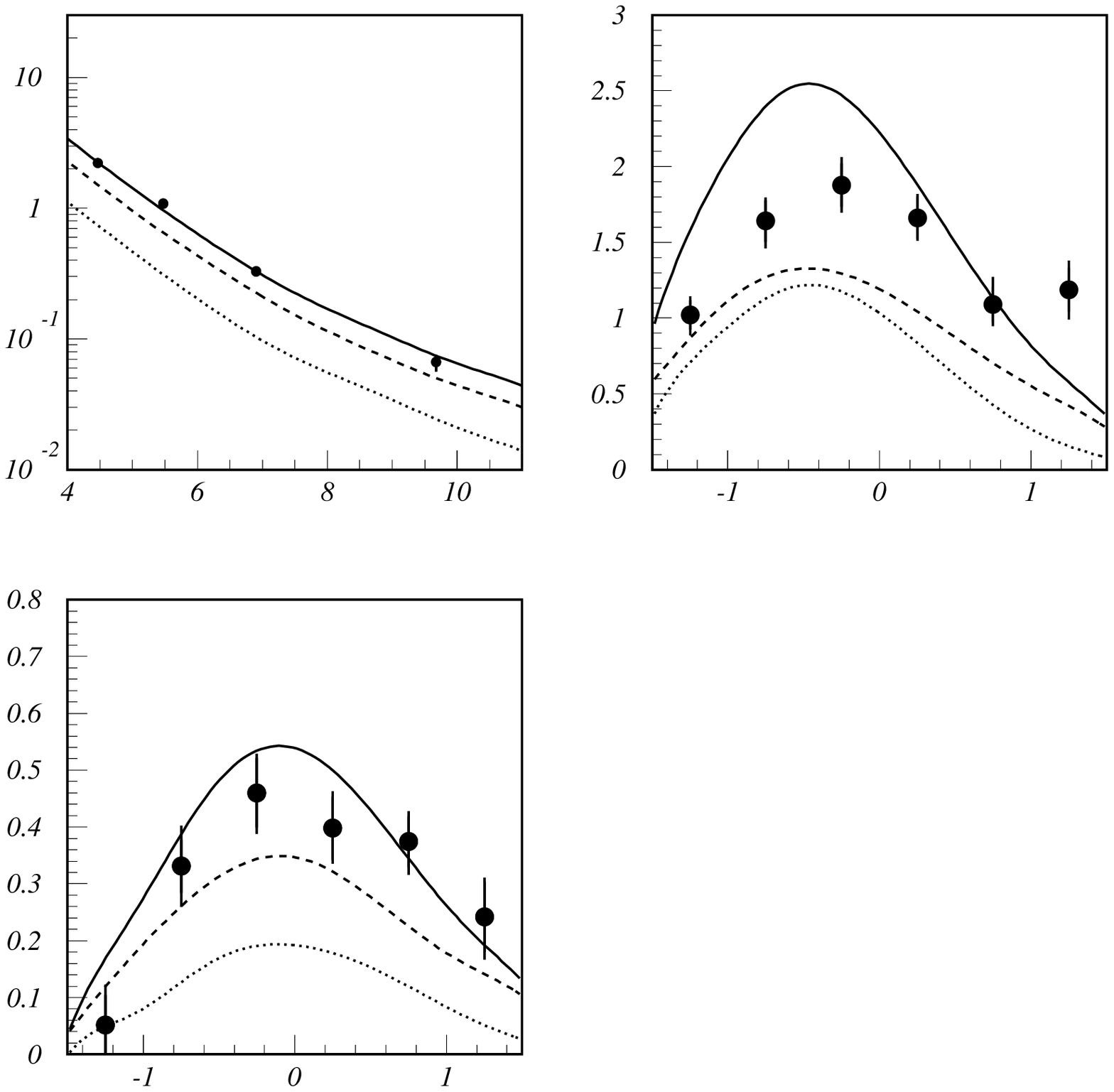}
\begin{picture}(450,450)\put(-10,350){$d\sigma/dp_T$, nb/GeV}
\put(240,350){$d\sigma/d\eta$, nb}
\put(-10,100){$d\sigma/d\eta$, nb}
%\put(240,100){$d\sigma/d\eta$, nb}
\put(150,120){$p_T$, GeV}
\put(400,120){$\eta$}\put(150,-130){$\eta$}\put(345,300){$p_T>4$ GeV}
\put(95,50){$p_T>6$ GeV}
\put(-40,-180){Fig.~3. \parbox[t]{14cm}{
The experimental distributions of the $D^*$-mesons photoproduction in
transverse momentum ($p_T$) and pseudorapidity ($\eta$) for
$130<W<280 \ {\rm GeV}$ and $Q^2<1\ {\rm GeV^2}$ in  comparison
with the BKL predictions \cite{BKL} (dashed curve), with the VDM
prediction(dotted curve)  and  with  the sum of BKL and VDM predictions
(solid curve).
}}
\end{picture}
\newpage
\vspace*{-3cm}\includegraphics{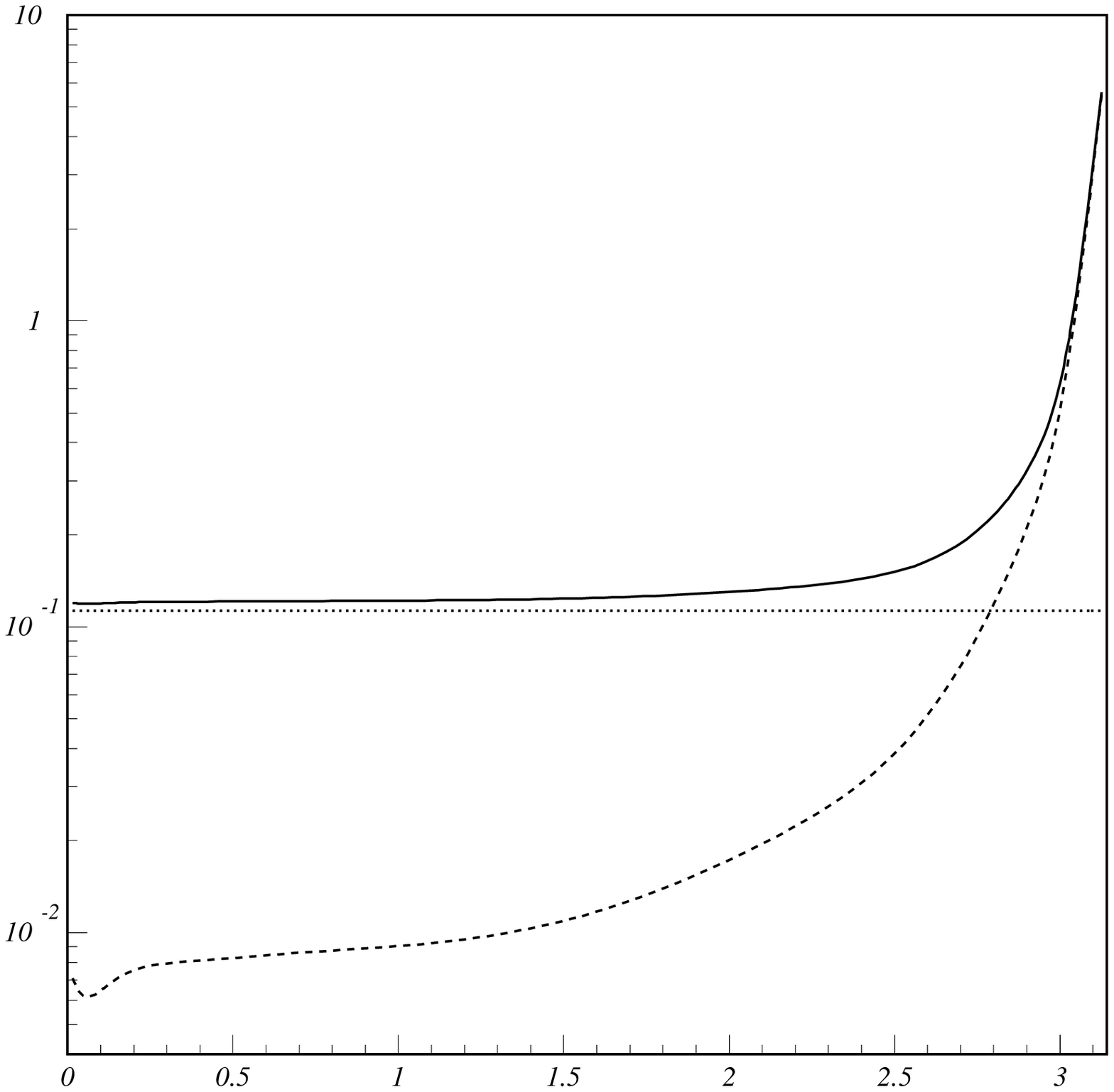}
\begin{picture}(450,450)\put(0,250){$d\sigma_{\gamma p}/d\Theta$, nb}
\put(360,-130){$\Theta$}
\put(20,-180){Fig.~4. \parbox[t]{12cm}{
The azimuthal correlation between  two charmed particles
for the $D^*$-meson photoproduction in the kinematical region of
$p_T>6$~GeV (dashed curve is the BKL contribution \cite{BKL},
dotted curve is the VDM contribution, solid curve  is the
sum of BKL and VDM prediction). $\Theta$ is the angle between the transverse
momenta of the charmed particles.
}}
\end{picture}

\end{document}